\documentclass[lettersize,journal]{IEEEtran}
\usepackage{amsmath,amsfonts}
\usepackage{algorithmic}
\usepackage{algorithm}
\usepackage{array}
\usepackage[caption=false,font=normalsize,labelfont=sf,textfont=sf]{subfig}
\usepackage{textcomp}
\usepackage{stfloats}
\usepackage{url}
\usepackage{verbatim}
\usepackage{graphicx}
\usepackage{cite}
\usepackage{booktabs}
\usepackage{makecell}
\usepackage[T1]{fontenc}
\usepackage{xcolor}
\usepackage{soul}
\sethlcolor{yellow}
\newcommand{\rev}[1]{{#1}}
\begin{document}

\title{Leveraging fNIRS to Evaluate Workload for Adaptive Training in Virtual Reality}

\author{Cara A. Spencer, Christopher D. Wickens, Jalynn B. Nicoly, James Crum, Benjamin A. Clegg, Joanna E. Lewis, Francisco R. Ortega, Lucas Plabst, Rebecca L. Pharmer, and Leanne Hirshfield
\thanks{Manuscript received July 24, 2025; revised May 5, 2026.}
\thanks{This research was supported by Natalie Steinhauser and the Office of Naval 
Research conducted under grant numbers ONR N00014-24-1-2214 and ONR 
N00014-23-1-2298. Corresponding author: Cara A. Spencer. All authors were 
responsible for the design of the study and methodology, and revising the 
manuscipt. CAS, CDW, JBN, JC and LH were responsible for data acquisition, 
analysis and interpretation, as well as drafting and revising the manuscript.}
\thanks{Cara A. Spencer, Jalynn B. Nicoly, James Crum and Leanne Hirshfield are from the University of Colorado Boulder, CO, USA (emails: cara.spencer@colorado.edu, jalynn.nicoly@colorado.edu, james.crum@colorado.edu, and  leanne.hirshfield@colorado.edu).}
\thanks{Christopher D. Wickens, Francisco R. Ortega, Lucas Plabst and Rebecca L. Pharmer are with Colorado State University, CO, USA (emails: chris.wickens@colostate.edu, fortega@colostate.edu,  Lucas.Plabst@colostate.edu, rebecca.pharmer@colostate.edu).}
\thanks{Benjamin A. Clegg is with Montana State University, MT, USA (email: benjamin.clegg@montana.edu).}
\thanks{Joanna E. Lewis is with the University of Northern Colorado, CO, USA (email: Joanna.lewis@unco.edu).}
\thanks{This work involved human subjects or animals in its research. Approval of all ethical and experimental procedures and protocols was granted by the Ethics Review Board at the University of Colorado Boulder and performed in line with university requirements.}
\thanks{This article has supplementary material provided by the authors and color versions of one or more figures available at http://ieeexplore.ieee.org}}

\markboth{Manuscript ID: }%
{Shell \MakeLowercase{\textit{et al.}}: A Sample Article Using IEEEtran.cls for IEEE Journals}


\maketitle

\begin{abstract}
Advances in technology offer the potential for future adoption of a combination of virtual reality (VR) and real-time adaptivity to enhance training and education. Providing a valid neuro-ergonomic measure of cognitive load can enable an adaptive training regime to continuously adjust task difficulty to an optimal level as training progresses. The current study validated the functional near-infrared spectroscopy (fNIRS) measure of cognitive load to reflect the demands of two different forms of load within Cognitive Load Theory: extraneous and intrinsic to the task to be mastered. Thirty-six participants completed a VR shape assembly training task followed by a test of their skill retention. They wore near–full head coverage fNIRS and provided subjective ratings of their workload. The fNIRS findings largely corroborate intrinsic workload literature with significant activation in cortical regions (dorsolateral and rostral prefrontal cortex and left angular gyrus) associated with working memory, short term memory buffers, multisensory integration, and attention. These fNIRS results were tracked closely by NASA TLX measures of mental workload. The results also revealed far less brain activity associated with extraneous load, namely just the right angular gyrus, deemed irrelevant to the mastery of the task.
\end{abstract}


\section{Introduction}
\IEEEPARstart{C}{omputer}-based alternative learning environments, like virtual reality (VR) and augmented reality (AR), facilitate versatile and controlled training where various stages of learning can be manipulated, evaluated, and optimized \cite{ref34}. Training refers to an information delivery program for the systematic acquisition of skills \cite{ref11}. When well-designed, these alternative learning environments help overcome training limitations, including high cost, setup time, facility access, ease of instruction, and safety risks \cite{ref65}. Moreover, VR can leverage a more ecologically plausible environment for training increasing embodiment and immersion \cite{ref35}. Despite this potential, using VR comes with its own challenges. For example, Shinde et al. \cite{ref51} reported that while participants rated the perceived usefulness of VR high, they also reported as challenging or burdensome to use. Crucially, the need for more systematic, theory-driven approaches to the application of AR and VR in training can be seen from the absence of a general advantage from current uses of these tools \cite{ref29}. The potential benefit of using VR in learning motivates researchers to seek to mitigate these challenges.

The research community has noted that the success of many adaptive training paradigms for procedural learning can be attributed to being firmly grounded in the well-validated Cognitive Load Theory (CLT) of instructions \cite{ref54, ref55, ref40, ref56, ref63}. According to CLT, at any point in a learning sequence, three sources of cognitive workload impose on the learner. Extraneous load refers to the processing effort imposed by non-task-relevant aspects of the learning environment (e.g., poor text font, poorly worded or wordy instructions). Intrinsic load is essential processing effort associated with the inherent characteristics of the task to be trained (e.g., flying a helicopter has higher intrinsic load than driving a car). It is often characterized by working memory demands or number of task component interactions. Germane load is the cognitive workload imposed as the learner invokes strategies to master the task itself and transfer its elements into long term memory. This load is imposed by elements such as rehearsing task components, “chunking” its multiple elements into a smaller number, or imposing self-testing \cite{ref44}.  

Traditionally, three categories of techniques have been used to measure the resources demanded by and deployed for task performance (i.e., the combination of intrinsic and extraneous load) and hence, inversely, the resources available for learning (i.e., germane load). These include (1) performance on a secondary task done concurrently with the primary task to be learned where lower secondary task performance indicates higher workload on the primary task; (2) subjective workload ratings \cite{ref23, ref43, ref28}; and (3) physiological or neuro-ergonomic measures that either correlate with workload (e.g., heart rate variability), or directly assess brain activation corresponding to resource demand (e.g., electric potentials from neural activation or hemodynamics in the brain \cite{ref4}). The high likelihood for interference with learning dissuades the use of either of the first two types of techniques for concurrently indexing workload in real-world training. 	Recent advances in brain-computer interfaces, with new wearable and non-invasive brain measurement devices, however, offer the potential to improve adaptive training in complex VR environments. The current study thus investigates the use of the non-invasive functional near-infrared spectroscopy (fNIRS) brain measurement modality, a device particularly well-suited for assessment of brain activation in real-world and VR/AR settings \cite{ref26}. We infer changes in intrinsic load and extraneous load while participants conducted a shape assembly task in VR, paving a pathway toward future adaptive training platforms in VR, based on objective fNIRS measurements. 

In the following sections, we describe related literature in Cognitive Load Theory (CLT), adaptive training, and the use of non-invasive brain measurement to assess cognitive load. We then describe research challenges in current state of the art measurements of cognitive load and we propose the use of fNIRS to provide an objective and diagnostic measure of cognitive load during training in VR.  We pose four hypotheses, and we describe an experiment carried out to investigate those hypotheses, in order to demonstrate the utility of fNIRS for assessing cognitive load during training in VR.  In our discussion we describe our findings, which suggest that fNIRS is a promising modality for measuring cognitive load in real-time for adaptive training in VR.

\subsection{Adaptive Training}
A key to efficient learning from CLT literature is to adaptively increase the intrinsic load of the task as learning progresses to preserve an adequate amount of mental resources available to allocate to germane load \cite{ref32}. Extraneous load must also be minimized since it accelerates the depletion of available mental resources. This depletion constrains overall processing capacity inhibiting intrinsic and germane load, which thus constrains the ability to transfer skills learned to new applications \cite{ref53}. Stated in other terms, a goal of adaptive training is to maintain a desirable (level of) difficulty \cite{ref6}. Recent work on adaptive VR training has shown that adaptively maintaining this desired difficulty has a positive impact on learning and subjective experience of workload \cite{ref61}.

\subsection{Brain measurement of workload}
\rev{In recent decades, several methods have been used for measuring latent states during VR tasks, including psychometric, behavioral, neuroimaging, and physiological methods. These research efforts have been supported by VR devices beginning to incorporate sensors like eye-tracking, accelerometers, and heart rate monitors into the hardware for mental state estimation. Neuroimaging techniques like fNIRS and electroencephalography (EEG) are frequently used for more direct insight because they capture neural activity and are often portable and adaptable} \cite{ref33, ref66}. Aksoy et al. \cite{ref2}, \rev{for example, collected concurrent fNIRS and behavioral performance during VR-based medical learning modules and found the measures to be complementary for an accurate assessment of learning. Others have also found integration of sensors creates opportunities for better mental state estimation through the combination of several signals} \cite{ref18, ref42}. \rev{Integrating these metrics within VR environments (e.g., spanning pedagogical simulations to therapeutic interventions) affords a high-fidelity lens into latent cognitive workload dynamics that traditional behavioral outputs may fail to capture} \cite{ref22, ref31}.

\rev{Generally, within training and learning neuroimaging  studies the prefrontal cortex (PFC) is the most common region of interest, followed by the anterior cingulate cortex and posterior parietal cortex since all three have repeatedly shown sensitivity to changes in workload (depicted below in Fig. 1)} \cite{ref68, ref3}. \rev{However, research in complex VR domains, such as driving instruction, has successfully expanded this scope to include non-frontal regions, demonstrating that temporo-occipital activation also serves as a robust predictor of workload dynamics} \cite{ref46, ref59}. \rev{This expansion toward a whole-brain perspective aligns with network-level research that has identified the frontoparietal and cingulo-opercular networks as playing roles in the cognitive control of workload and multisensory integration} \cite{ref49, ref62}.  \rev{The frontoparietal network is especially influential during the early skill acquisition phase of learning due to its role in managing working memory and effortful reconfiguring of mental models by adapting to feedback and integrating novel multisensory stimuli} \cite{ref24, ref57}. \rev{This network works synergistically with the cingulo-opercular network to monitor errors and sustain task maintenance and goals across stages of learning and during the performance of that skill} \cite{ref64}.

\begin{figure}[!t]
\centering
\includegraphics[width=2.5in]{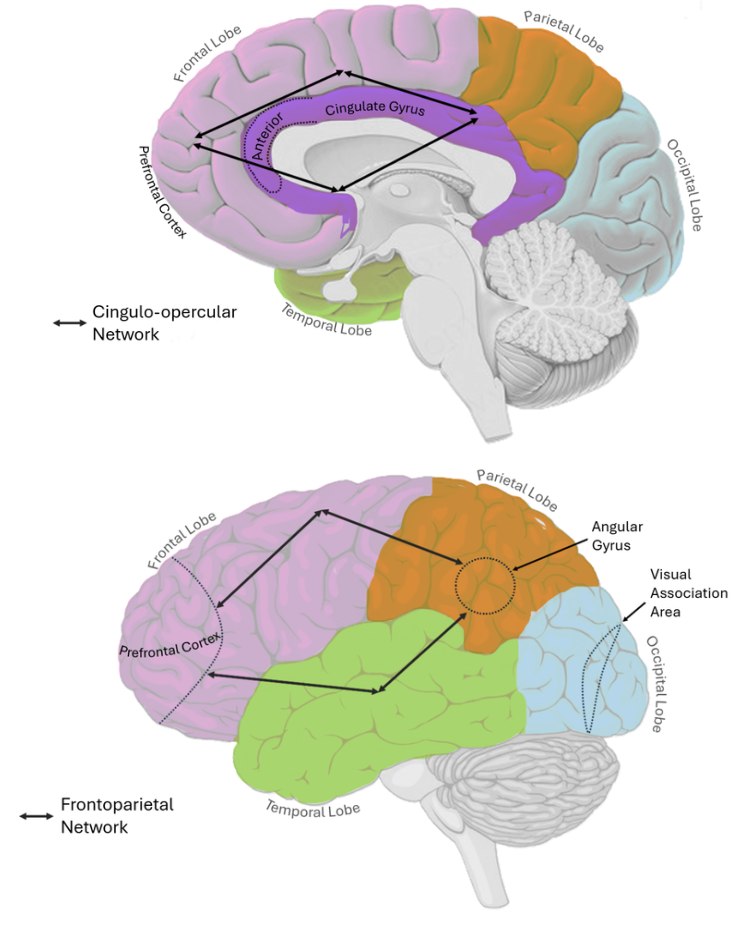}
\caption{Brain regions and networks implicated in workload and training studies. The top image shows a lateral view of the left hemisphere. Each lobe is color-coded where pink is the frontal lobe, orange is the parietal lobe, blue is the occipital lobe, and green is the temporal lobe. Three regions from prior work are highlighted with dotted lines namely: PFC, Angular Gyrus, and Visual Association Cortex. A high level representation of the frontoparietal network is depicted with solid two-way arrows. The bottom image shows a sagittal view of the left hemisphere. Each lobe is color coded the same way. Additionally, the cingulate gyrus is colored purple. The main region from prior work highlighted with the dotted line is the anterior portion of the cingulate gyrus. The cingulo-opercular network is likewise depicted with solid two-way arrows.}
\label{fig_1}
\end{figure}

\rev{As previously established, regions along these networks have been shown to be sensitive to workload, and this is for several reasons. The PFC directly manages the active manipulation of complex task elements and top-down attention: encoding and maintaining active representations and goals, selecting relevant information and actions, and suppressing irrelevant information and actions} \cite{ref38}. \rev{Because the PFC acts as a central bottleneck for information processing, its capacity to provide these biasing signals is inherently limited, making it a sensitive indicator of workload; as task complexity increases, current theories claim the metabolic demand on PFC hierarchical circuits rises until a ceiling is reached in the capacity to manage information} \cite{ref30}.

\rev{In conjunction, the angular gyrus (AG) in the posterior parietal regions of both hemispheres acts as a major connection and routing hub for the integration of sensory information by dynamically combining semantic and episodic memory during sense making} \cite{ref25, ref50}. \rev{While the AG is traditionally identified as a hub within the task-negative Default Mode Network (DMN), it functions as a critical multimodal convergence zone that can flexibly shift its connectivity between the DMN and the FPN to support task-specific demands} \cite{ref60}. \rev{While the AG in both hemispheres support multisensory integration, the left AG is more sensitive to the semantic, linguistic, and rule-based complexities intrinsic to the task and the right AG is preferentially recruited for the spatial reorienting and attentional filtering required to manage environmental noise} \cite{ref52, ref50}. \rev{The regions support the understanding of spatial relationships in the environment, retrieval of spatial information and expectancies, and the mediation of language in making sense of the current context} \cite{ref7, ref50}. \rev{Due to its role in information integration, the AG operates under the top-down modulation of the PFC; consequently, as executive demands increase, the AG’s capacity for information integration is constrained by the same bottleneck of resources that governs the broader frontoparietal network} \cite{ref38, ref30}. 
 
In terms of measurement of activity in these regions, fNIRS is useful for capturing mental states that emerge over a longer timescale like workload since it records changes in the slow hemodynamics of the brain (e.g., timescale of seconds, compared to EEG’s milliseconds sampling rate). Hirshfield et al. \cite{ref26} carried out an extensive review of 54 fNIRS studies of cognitive load. This review revealed that only a small portion of these (N=14), measured workload in complex tasks with brain measurements made in cortical regions beyond just the PFC region (which is under the forehead, the most common region for fNIRS measurements to be made). Crucially, none of the generalizable studies evaluated fNIRS-based workload metrics within the high-fidelity constraints of a VR training environment, representing a significant gap in the literature that the present study aims to address.  
 
\rev{In addition to their review, Hirshfield et al.} \cite{ref26} \rev{conducted a study and found fNIRS is both sensitive and diagnostic to cognitive load types and levels in complex tasks. This study employed a perceptual-motor task with substantial cognitive (working memory) requirements, to reveal the close correspondence of fNIRS measures to both secondary task and subjective measures of cognitive load. Specifically, decreases in deoxygenated blood (HbR), which indicate increased neural activation, was associated with working memory load in the PFC by region of interest (ROI) and channel level analysis techniques. Additionally, this study differentiated working memory load from visual load revealing that visual load produced elevated oxygenated blood (HbO) in the left AG. This did not overlap with channels activated by working memory load: elevated HbO in the visual association cortex (left superior occipital gyrus) and decreased HbR in the left ventrolateral PFC (inferior frontal gyrus), right primary motor cortex (precentral gyrus) and right AG.}  
    
While it is well established that cerebral blood flow results from higher cognitive load and can be effectively measured by fNIRS in realistic tasks, it is also important to go beyond this. The relationships between the types and levels of workload evoked by skill mastery and brain regions activated still needs examining. In particular, differentiating those that may be intrinsic to the task, those that are distracting from the learning process (extraneous load) and those that are directly associated with skill mastery (germane load), is of crucial importance to measure for adaptive training. Traditional adaptive training procedures measure the learner’s task performance (e.g., accuracy) to infer proficiency and adjust difficulty. However, relying on task performance may not fully capture the learner’s cognitive state hindering the ability to maintain the desired level of difficulty. Incorporating direct measures of cognitive workload from neural signals may provide a more accurate assessment. For example, functional near-infrared spectroscopy (fNIRS) has been used for adapting difficulty based on workload estimation specifically during a path planning scenario \cite{ref1}. fNIRS is a non-invasive optical neuroimaging technique that quantifies cortical hemodynamic responses—specifically changes in oxygenated and deoxygenated hemoglobin concentrations—as an indirect measure of neuronal metabolic activity.

\subsection{Research Contribution and Hypotheses}
Our experiment aims to address gaps outlined above by implementing a study in VR where participants learned how to assemble complex shapes under two different manipulations of cognitive load. The study included two phases, training and retention testing, to assess the effect of training workload on the transfer of assembly skills. We varied intrinsic load by manipulating the task’s working memory demands. We varied extraneous load by manipulating the amount of redundant text in the instructions (i.e., wordiness). Participants were trained to assemble four shapes, with training which had attributes of high or low intrinsic load and high or low extraneous load. The adaptivity of the program was implemented by removing training instructions based on accuracy in order to force reliance on memory. In addition to the fNIRS measures, we collected error rate, completion time, and self-report NASA-Task Load Index (TLX) workload values throughout the task. We designed a neuro-sensor montage to cover specific brain regions such that the fNIRS sensors were positioned to measure the frontal and parietal lobes, with some coverage in temporal and occipital areas as well (full spatial coverage is detailed in the Methods). Building on prior fNIRS training research for measuring cognitive load and incorporating findings from the very relevant study by Hirshfield et al. \cite{ref26}, we propose four hypotheses grounded in CLT. 

\textbf{H1}: fNIRS mis a measure of intrinsic load: We posit that, during training, the fNIRS will be sensitive to intrinsic load, and that brain regions involved in executive function and working memory will show significant increases in activation when intrinsic load is increased. These brain regions are hypothesized to be in the PFC and the angular gyri. 

\textbf{H2}: Further, these increases will align with our NASA TLX, training time and error rate which can serve as manipulation checks in our study. 

\textbf{H3}: fNIRS is a measure of extraneous load: We posit that, during training, the fNIRS will be sensitive to increases in extraneous load as we varied it here, and that brain regions involved in reading, visuospatial attention, and multisensory integration will show significant increases in activation when extraneous load is increased. Specifically, we expect activation in the angular gyri. 

\textbf{H4}: Further, these increases will align with our NASA-TLX, training completion time, and error rate metrics, which can serve as manipulation checks in the study. 

\section{Method}
We designed and implemented an experiment to investigate our hypotheses. During the within-subjects experiment participants completed a shape assembly task in VR while wearing fNIRS.

\begin{figure*}[!t]
\centering
\includegraphics[width=5in]{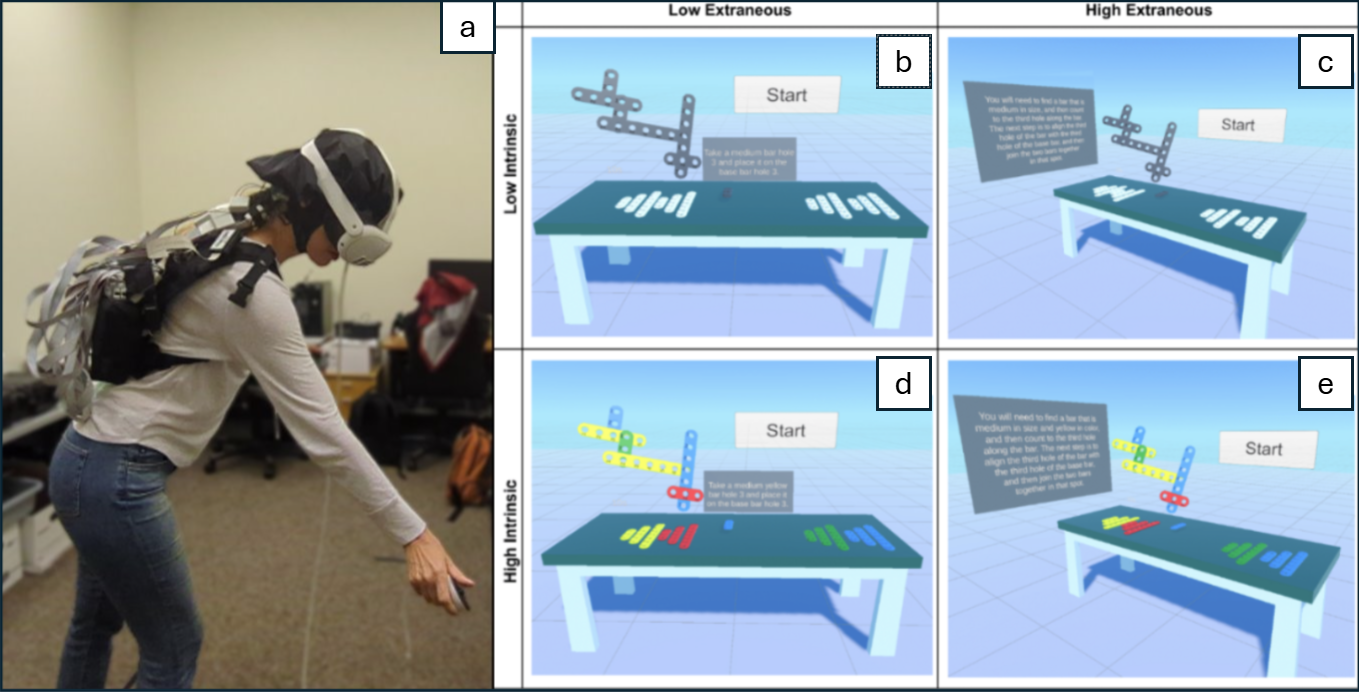}
\caption{(a) A participant wearing the fNIRS and VR Head-Mounted Display during the experiment. (b) The low extraneous, low intrinsic condition where all bars are the same color and the instructions are simple. (c) the high extraneous, low intrinsic condition where the bars are the same color but the instructions are verbose and off centered. (d) The low extraneous, high intrinsic condition where the bars are different colors and instructions are simple. (e) The high extraneous, high intrinsic condition where the bars are different colors and the instructions are verbose and off-centered.}
\label{fig_2}
\end{figure*}

\subsection{Participants}
The study recruited 36 participants, ranging in age from 18 to 49 (Mean = 24 years, SD = 9.95, \rev{Median = 22, IQR = 3}). The sample consisted of 47\% females (N = 17) and 53\% males (N = 19), with no participants identifying as non-binary. \rev{Five reported being left-handed and the rest reported right-handedness.} Ten participants indicated prior experience with AR, while 34 reported previous VR use (M = 0.88 hours per week, SD = 3.52).  \rev{All participants either had a bachelor's level of education or were in the process of attaining a bachelor's degree. Participants were recruited using flyers at a University in the western United States. Inclusion criteria included no reported VR cyber sickness and either correct or perfect vision.} An a priori power analysis of a medium effect size, with alpha set at .05 and 1-beta (power) at .80, determined that N = 34 was adequate for a within-subjects comparison of workload. All participants were compensated monetarily (\$40). Recruitment and experimental procedures were approved by the University's Institutional Review Board. 

\subsection{\rev{Protocol}} 
The experimental protocol consisted of four rounds in VR, each involving learning how to assemble a unique shape (see task flow diagram in the Supplementary Materials). Four unique shapes were composed of bars of varying lengths and colors connected at specific attachment points (Fig. 2). To mitigate order effects, the presentation of shapes was counterbalanced across sessions. Before the experimental rounds, participants completed a practice assembly of a shape to familiarize themselves with the VR control interface. Each round followed a five-phase sequence: (1) \textit{training}, (2) \textit{distraction}, (3) \textit{retention test}, (4) \textit{workload assessment}, and (5) \textit{rest}.

During the \textit{training} phase, participants performed six shape assembly repetitions using a fading instruction protocol. During the first assembly, the VR system provided step-by-step sequential instructions for all five assembly steps. In each subsequent assembly repetition, one instruction was removed and replaced with the prompt "Please perform Step X," where X is the current step, requiring participants to recall the step from memory. By the sixth repetition, each step instruction was replaced by the prompt “Please perform Step X.” The system provided immediate feedback; a green checkmark for correct placement or a red “X” for errors that were shown until the step was performed correctly.

There were two manipulations implemented in the study: Intrinsic Load (high/low) and Extraneous Load (high/low), as shown in Fig. 2. Extraneous load was manipulated by varying the instructional text verbosity while keeping the quantifiable information content the same. The instructional text was also offset to the side spatially adding additional effort to read it. In the low extraneous load conditions, instructions were concise, such as: “Take a medium red bar hole 3 and place it on the base bar hole 3.” In the high extraneous load conditions, instructions were more indirect, such as: “You will need to find a bar that is medium in size and red in color and then count to the third hole along the bar. The next step is to align the third hole of the bar with the third hole of the base bar and then join the two bars together in that spot.” Whereas intrinsic load was characterized by the number of attributes or “rules” that participants had to recall. In the low intrinsic load conditions, the two attributes were bar length and attachment point. In the high intrinsic load conditions, color was included as an additional attribute. 
     
Immediately following training, participants performed a 90-second \textit{distraction} task to prevent active rehearsal. This task required counting the bounces of a virtual ball. In the subsequent \textit{retention test}, participants were instructed to build the shape they just learned entirely from memory without instructional cues or performance feedback. Following the retention test, the experimenter verbally administered a \textit{workload assessment}: the mental and physical demand subscales of the NASA Task Load Index (NASA-TLX) \cite{ref23}. Each round concluded with a 15-second \textit{rest} period before proceeding to the first assembly of the next shape. The VR platform automatically logged all behavioral metrics, including reaction times, completion times, and the frequency and types of errors. 

\subsection{fNIRS Setup}
Participants were donned with the fNIRS cap, a black shower cap to reduce ambient light, and the Meta Quest 3 head-mounted display. fNIRS signal acquisition of neural hemodynamics was acquired using two NIRSport2 devices (NIRx) that employ continuous-wave illumination at two different wavelengths (760nm and 850nm) at 3.81 Hz. A custom montage of fNIRS channels was designed (see Supplementary Materials for figure). The configuration of source and detector optodes formed 80 channels, with 8 short-separation channels to account for extracerebral noise. This configuration achieved spatial coverage over most of the frontal and parietal lobes, with some coverage in temporal and occipital areas. Anatomical locations of each optode were determined using a Patriot 3D Digitizer (Polhemus) using five fiducial landmarks including inion, nasion, right tragi, left tragi, and top center (Cz). Subject-specific spatial coordinates from one participant were used as a template for the remaining sample. These digitized spatial data were registered to a standard Montreal Neurological Institute (MNI) space using linear interpolation (e.g., MNI-ICBM152 \cite{ref37}), which allows for greater spatial comparability with other neuroimaging methods. These coordinates were applied to a brain atlas for anatomical labeling (NIRS-SPM \cite{ref67}) in MATLAB (MathWorks, Inc.), yielding Brodmann’s areas (BA) for each channel. The full list of channels, coordinates, and anatomical regions is in the Supplementary Materials.

\section{results}
\subsection{Load Manipulation Check Using Behavioral and Self-Report Data }
Before proceeding with analyses of neural signals, we first ensured cognitive workload was indeed manipulated by our experimental conditions via traditional metrics: training and retention test completion time, training and retention test error rates, and self-report cognitive load. Completion time was evaluated as the average time taken to complete the first six trials prior to the distractor task. Error rate was evaluated as the percentage of errors in shape construction (e.g., incorrect color, size, or placement). 

Results suggested that our manipulation of intrinsic load was indeed effective, specifically results showed that high intrinsic load significantly increased training and testing completion time (M = 283.25, SD = 83.67; M = 31.43, SD = 12.23), training error rate (M = 15.06\%, SD = 9.79), and the averaged mental and physical workload ratings (M =  10.11, SD = 4.53; M = 5.00, SD = 4.06). The effects of high extraneous load were less robust with the high load condition only significantly increasing training completion time (M = 294.07, SD = 95.88). These findings support \textbf{H2} and partially support \textbf{H4} that these manipulations will influence workload and the behavioral performance measures. See Supplementary Materials for figures and Table I for a summary of the results. 

\begin{table*}[!t]
\caption{Mean (M) and standard deviation (SD) of completion time (seconds), error rate, and workload ratings across all four conditions.}
\label{tab_descriptive}
\centering
\footnotesize
\setlength{\tabcolsep}{3pt}
\renewcommand{\arraystretch}{1.1}
\begin{tabular}{l cc cc cc}
\toprule
 & \multicolumn{2}{c}{Completion Time} & \multicolumn{2}{c}{Error Rate} & & \\
\cmidrule(lr){2-3} \cmidrule(lr){4-5}
Condition & Training & Retention Test & Training & Retention Test & Mental Load & Physical Load \\
\midrule
Low Intrinsic, Low Extraneous   & \makecell{M = 224.00 \\ SD = 90.40}  & \makecell{M = 23.14 \\ SD = 8.58}  & \makecell{M = 12.27\% \\ SD = 7.83}  & \makecell{M = 44.29\% \\ SD = 37.85} & \makecell{M = 9.32 \\ SD = 4.37}  & \makecell{M = 3.43 \\ SD = 3.34} \\
Low Intrinsic, High Extraneous  & \makecell{M = 294.07 \\ SD = 95.88}  & \makecell{M = 28.43 \\ SD = 14.90} & \makecell{M = 13.64\% \\ SD = 10.26} & \makecell{M = 42.86\% \\ SD = 27.60} & \makecell{M = 9.89 \\ SD = 4.75}  & \makecell{M = 4.00 \\ SD = 3.82} \\
High Intrinsic, Low Extraneous  & \makecell{M = 283.25 \\ SD = 83.67}  & \makecell{M = 31.43 \\ SD = 12.23} & \makecell{M = 15.06\% \\ SD = 9.79}  & \makecell{M = 33.57\% \\ SD = 30.82} & \makecell{M = 10.11 \\ SD = 4.53} & \makecell{M = 5.00 \\ SD = 4.06} \\
High Intrinsic, High Extraneous & \makecell{M = 367.82 \\ SD = 120.39} & \makecell{M = 31.71 \\ SD = 14.69} & \makecell{M = 17.69\% \\ SD = 8.75}  & \makecell{M = 37.14\% \\ SD = 37.60} & \makecell{M = 10.36 \\ SD = 4.57} & \makecell{M = 4.96 \\ SD = 4.83} \\
\bottomrule
\end{tabular}
\end{table*}

\subsection{Training Performance}
Training performance was assessed using completion time and error rate. A two-way, repeated-measures ANOVA revealed that both high intrinsic load ($F(1,27) = 36.78$, \textit{p} $< .01$, $\eta^2 = .11$) and high extraneous load ($F(1,27) = 34.61$, \textit{p} $< .001$, $\eta^2 = .14$) increased training completion time with no significant interaction ($F(1,27) = 0.24$, \textit{p} $= .63$, $\eta^2 < .01$). Another two-way, repeated-measures ANOVA revealed that increasing intrinsic load significantly increased training error rate ($F(1,27) = 7.36$, \textit{p} $= .01$, $\eta^2 = .04$), while extraneous load had no significant effect ($F(1,27) = 1.81$, \textit{p} $= .19$, $\eta^2 = .01$). No significant interaction was observed between intrinsic and extraneous loads ($F(1,27) = 0.16$, \textit{p} $= .69$, $\eta^2 = .001$). 
\subsection{Retention Test Performance}
Retention test performance was also assessed using completion time and error rate. A two-way repeated measures ANOVA revealed a significant effect of intrinsic load ($F(1,27) = 10.33$, \textit{p} $= .003$, $\eta^2 = .05$) lengthening retention test completion time at higher load. Extraneous load had no significant effect ($F(1,27) = 2.48$, \textit{p} $= .13$, $\eta^2 = .01$). No significant interaction was observed between intrinsic and extraneous loads ($F(1,27) = 1.70$, \textit{p} $= .20$, $\eta^2 = .01$). A two-way repeated measures ANOVA on error rate revealed no significant effect from either intrinsic load ($F(1,27) = 1.57$, \textit{p} $= .22$, $\eta^2 = .02$) or extraneous load ($F(1,27) = 0.06$, \textit{p} $= .81$, $\eta^2 = .00$). No significant interaction was observed between intrinsic and extraneous loads ($F(1,27) = 0.15$, \textit{p} $= .70$, $\eta^2 = .001$).
\subsection{Cognitive Workload}
Cognitive workload was assessed using the average mental and physical load ratings from the NASA-TLX on a 0 to 20 scale. \rev{The NASA-TLX is a standardized scale extensively utilized for subjective workload assessment in human-computer interaction studies} \cite{ref36}. Since workload was measured only once per shape, training and retention test phases were not separate. A two-way repeated measures ANOVA revealed, as hypothesized, significantly higher rated workload with higher intrinsic load ($F(1,27) = 11.82$, \textit{p} $= .002$, $\eta^2 = .02$). However, there was no significant effect of extraneous load ($F(1,27) = 1.71$, \textit{p} $= .20$, $\eta^2 = .003$). No significant interaction was observed ($F(1,27) = 0.55$, \textit{p} $= .46$, $\eta^2 = .001$).
\subsection{fNIRS Signal Processing}
The fNIRS channels indexed changes in concentrations of oxygenated ($\delta$ HbO2) and deoxygenated ($\delta$HbR) hemoglobin in the brain. More specifically, these local changes in concentrations of HbO and HbR in the brain were computed from optical densities using the modified Beer-Lambert Law \cite{ref13}. Changes in HbO and HbR signals were pre-processed to account for systemic sources of noise (i.e., physiological variability not due to task-related neural activity). This involved motion-artifact correction and filtering to correct serially correlated errors in the signals \cite{ref5}. Specifically, we employ optimal pre-whitening filters using autoregressive models and iteratively reweighted least squares (AR-IRLS). This is done by convolving a hemodynamic response function with stimulus onsets and durations from the behavioral data, resulting in GLM-based beta estimates for each channel in the single-subject design matrices. \rev{Because the data were collected only from the ROIs, which were specified a priori, in that there were no whole-brain contrasts, corrections for multiple comparisons were not applied to the results.} Data from other sensors were also included as regressors in these matrices to further account for sources of noise, namely the short-separation channels and the accelerometer data. Several datasets had to be discarded due to issues with the data quality or with the trigger software, resulting in 23 participant datasets for fNIRS analyses.

The second-level analysis consisted of random-effects analyses via summary statistics and contrasts \cite{ref15}. Group-level contrast effects were normalized to standard MNI space using linear transformations and interpolated onto a 3D brain mesh for visualization. Lastly, all analyses were conducted on both HbO and HbR, but the interpretation of results was based on research suggesting that HbR signals are less affected by systemic confounds \cite{ref14}. For example, fNIRS paradigms involving large bodily movements (e.g., moving arms in VR) or overt speech produce changes in arterial CO2 that—likely due to changes in respiration—alter the HbO signal to a greater degree than HbR \cite{ref47, ref48}.  

\subsection{fNIRS Results}
To investigate \textbf{H1}, within-brain statistical comparisons for intrinsic load (High > Low) during training showed significant increases in activation in several regions (uncorrected). Results are overlaid over the brain in Fig. 3 and listed in Table 2 (see Supplementary Materials for full channel list). Specifically, associated with higher intrinsic load, there were significant decreases in HbR in left rostral PFC (BA10; Channel 3), $t(22) = -3.09$, \textit{p} $< .01$, left dorsolateral PFC (BA46; Channel 15), $t(22) = -2.12$, \textit{p} $= .04$, left angular gyrus (BA39; Channel 40), $t(22) = -2.44$, \textit{p} $= .02$, and pre-motor and supplementary motor cortices (BA6; Channels 65 \& 76), with greatest activation changes in left BA6, $t(22) = -3.28$, \textit{p} $< .01$.

To investigate \textbf{H3}, within-brain statistical comparisons for extraneous load (High > Low) during training showed significant increases in activation in a single region, namely right angular gyrus (BA39; Channel 73), $t(22) = -2.44$, \textit{p} $= .02$. Inferential statistics for these contrasts are also provided in Table 2, with results overlaid over the brain in Fig. 4. 
\begin{figure}[!t]
\centering
\includegraphics[width=3.5in]{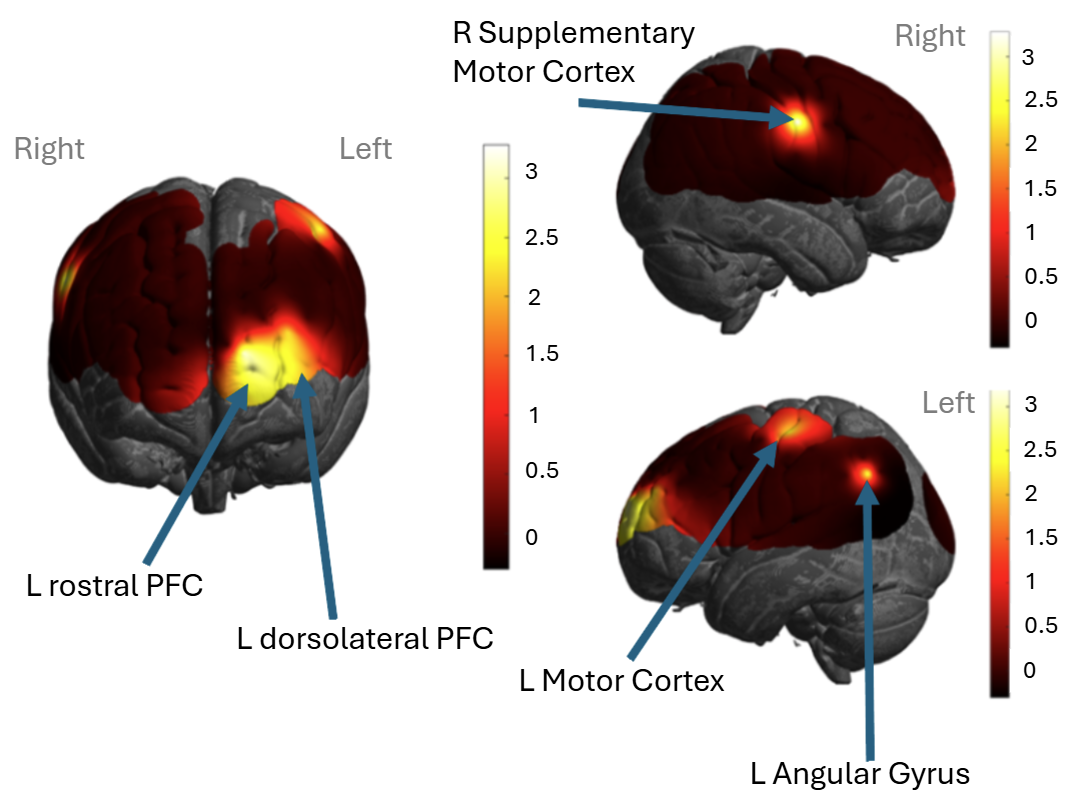}
\caption{Intrinsic load: High > Low (n = 23, p < .05, HbR). Greater changes in activation as a result of increasing intrinsic load are represented in white and yellow (i.e., larger t-values) with little to no changes in activation represented in red and black, respectively.  The left figure depicts the front of the brain. The two figures on the right represents the brain viewed from the right (top) and left (bottom) side. This figure depicts the activated areas for this condition: left rostral PFC, left dorsolateral PFC, left angular gyrus and the pre-motor and supplementary motor cortices.}
\label{fig_3}
\end{figure}

\begin{figure}[!t]
\centering
\includegraphics[width=3in]{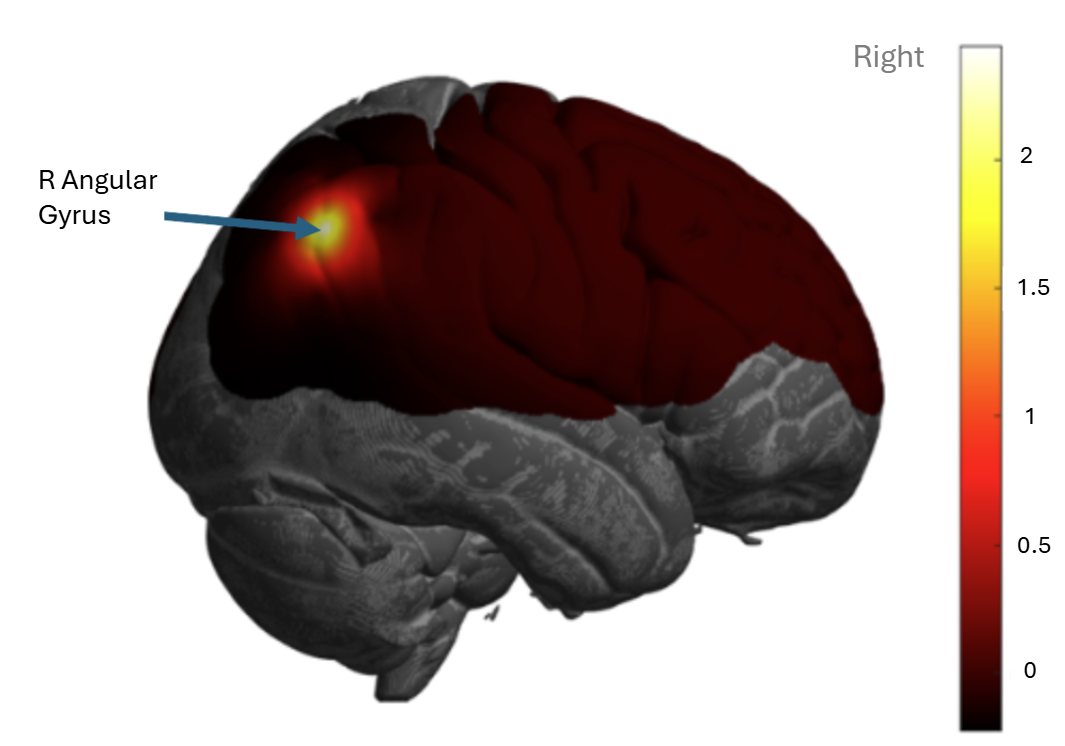}
\caption{Extraneous load: High > Low (n = 23, p < .05, HbR). Greater changes in activation as a result of increasing extraneous load are represented in white and yellow (i.e., larger t-values) with little to no changes in activation represented in red and black, respectively.  This figure depicts the right hemisphere of the brain and activated area for this condition: the right angular gyrus.}
\label{fig_4}
\end{figure}

\begin{table}[!t]
\caption{Channel-based Comparisons (HbR signals) of Intrinsic and Extraneous Load. For each condition (row), the channel number, beta value, t value, p value and Broadman Area (BA) are provided. Within the BA column, we also specify the functional label. The negative signs represent greater O2 absorbed upon return and hence greater activation at higher manipulated load.}
\label{tab_fnirs}
\centering
\footnotesize
\setlength{\tabcolsep}{4pt}
\renewcommand{\arraystretch}{1.2}
\begin{tabular}{p{0.5in} c r r c p{1in}}
\toprule
Condition Contrast & Channel & $B$ & $t$ & $p$ & BA \\
\midrule
Increasing Intrinsic   & 3  & $-27.92$ & $-3.09$ & $< .01$ & 10 -- Left frontopolar area \\
Increasing Intrinsic   & 15 & $-23.91$ & $-2.12$ & $.04$   & 46 -- Left dorsolateral prefrontal cortex \\
Increasing Intrinsic   & 40 & $-15.84$ & $-2.44$ & $.02$   & 39 -- Left angular gyrus, part of Wernicke's area \\
Increasing Intrinsic   & 65 & $-44.54$ & $-3.28$ & $< .01$ & 6 -- Right pre-motor and supplementary motor cortex \\
Increasing Intrinsic   & 76 & $-19.19$ & $-2.35$ & $.02$   & 6 -- Right pre-motor and supplementary motor cortex \\
Increasing Extraneous  & 73 & $-21.28$ & $-2.44$ & $.02$   & 39 -- Right angular gyrus \\
\bottomrule
\end{tabular}
\end{table}

\section{discussion}
The present study aimed to evaluate the impact of intrinsic and extraneous load during training, on retention test performance using a multimodal approach. We combined objective performance, subjective methods of workload, and neural activation to uncover relationships between cognitive workload, engagement, and learning. Our fNIRS results, as predicted, showed a difference in brain activity between load types (extraneous and intrinsic load) and within each load [manipulation (e.g. high – low intrinsic load and high – low extraneous load). High intrinsic load was identified as activating more brain areas than high extraneous load; this is consistent with Hirshfield et al. \cite{ref26} who found lower overall activation recorded by fNIRS when visual load increased than when working memory load was increased. From these between-condition comparisons across load manipulations, we can infer that manipulations imposed qualitatively differently loads to which our measures were differentially sensitive. Each of these findings can inform the development of a neuro-signature based adaptive training paradigm.    

\textbf{Hypothesis 1}: fNIRS and Intrinsic Load. With respect to intrinsic load, greater intrinsic load during training increased activation in anterior and posterior areas of the brain, particularly in the left hemisphere as depicted in Fig. 3. Specifically, there were greater hemodynamic changes in left dorsolateral PFC (BA46), left rostral PFC (BA10), and left angular gyrus (AG) (BA39) when intrinsic load was increased. This finding supports \textbf{H1} as increasing intrinsic load places relatively greater resource demands on the executive functions and working memory operations localized in the PFC, which should benefit learning. \rev{For example, the difference in stimulus complexity (i.e., colored bars) required additional visuospatial and relational information to be integrated, encoded, maintained, manipulated, and retrieved in combination with representations of other features of the stimuli during learning (i.e., size, number of holes, color, position, etc.). The AG is part of a multisensory integration area (inferior parietal lobule) that receives information from modality-specific sensory association areas (e.g., visual, auditory, somatosensory, and motor systems) to help make sense of experience} \cite{ref49}. \rev{Seeing a left lateralization, as opposed to the hypothesized right, here may indicate more mental resources allocated to the integration of the lexical-semantic information regarding the additional elements of the stimulus into procedural schema} \cite{ref50}. \rev{It suggests a reliance on verbal categorization strategies during the learning process (e.g., “attach the 3rd hole of the red bar to the 4th hole of the green”).}

More generalizable to a variety of training paradigms and building adaptive systems are the results from the left dorsolateral PFC and rostral PFC. \rev{This area is particularly important for encoding, manipulating, and updating information in short-term memory buffers} \cite{ref39, ref41}, \rev{as well as for response generation and inhibition} \cite{ref16, ref17}. \rev{These areas are involved in the frontoparietal network implicated in learning early in training (broadly depicted in Fig. 1)} \cite{ref24}. \rev{The observed activation changes in this area as a function of increasing intrinsic load are consistent with these functional specializations. In addition, rostral PFC supports attentional biasing (sustained and transient switching) between perception and stimulus-independent thought} \cite{ref8, ref9}. \rev{Such attentional biasing between modes of thought in the PFC is critical for successful multitasking during real-world behavior} \cite{ref10}. In the present work, because increasing intrinsic load likely affected aspects of encoding, manipulation, monitoring, and retrieval of information in working memory during learning, rostral PFC was plausibly recruited to sustain activation biasing in support of these cognitive operations. \rev{This area was likely also recruited for other stimulus-independent processes such as planning moves and self-initiated action (i.e., non-cued responses), potentially confirmed by the significant activation in the premotor and motor cortices.}     

When we consider how well the fNIRS results aligned with our other survey and behavioral measures, we note that the intrinsic load manipulation (i.e., defined by the additional attribute to remember) significantly increased training completion time and error rate, subject workload responses, and retention test completion time supporting \textbf{H2}. The finding that it did not reduce test performance documents the success of the adaptive training regime used here: the increased training time, reflecting more adaptive cycles, mitigated the effects of intrinsic load at retention.

\textbf{Hypothesis 3}: fNIRS and Extraneous Load. Greater extraneous load during training increased activation in a single area, namely right AG (rAG) as shown in Fig. 4. This is in line with the prediction posed by \textbf{H3}, that increasing extraneous load would place relatively greater resource demands on the integration of information (visuospatial, verbal, numeric, etc.), which should not benefit learning. \rev{The localized activation of the rAG, in the absence of broader recruitment within the canonical learning network, potentially serves as a signature of inefficient encoding driven by high extraneous cognitive load. This isolated recruitment, which deviates from our hypothesis and Hirshfield et al.} \cite{ref26} \rev{that both gyri would be involved, suggests that neural resources were disproportionately sequestered by the spatial split-attention effect; the rAG may have functioned as an attentional bottleneck for the constant bottom-up re-orienting between verbose instructions and the physical task} \cite{ref21}. \rev{Consequently, the brain may have been forced to rely on transient visuospatial working memory buffers to maintain unstable representations, preventing the transition of information into the left AG for semantic abstraction or the hippocampus for long-term integration} \cite{ref20}. \rev{This finding may also be attributed to the nature of this being a spatial learning task that is activating deeper brain regions fNIRS cannot reach like those found in fMRI spatial learning studies} \cite{ref27, ref12}. Taken together, the lack of left AG activation could indicate a failure to facilitate germane load, as the metabolic demands of navigating a load-eliciting interface inhibited the formation of a stable procedural schema.

When we consider how well the fNIRS results aligned with our other survey and behavioral measures, we note that the extraneous load manipulation only significantly increased training completion time (there was not significant effect on accuracy or NASA-TLX values or error rate), only lending partial supporting \textbf{H4}.

In summary, the significant activation of these brain regions when the intrinsic and extraneous load increased is consistent with previous research linking cognitive workload to brain hemodynamics (e.g., see Hirshfield et al. \cite{ref26}, for review). Therefore, these findings replicate and extend previous research using fNIRS to investigate cognitive workload. Importantly, results support the idea that this neuroimaging technique is markedly sensitive to quantitative changes within and between different types of workload during learning and task performance. In addition to sensitivity, the lack of PFC activation when extraneous load was increased and the presence of such activation when intrinsic load was increased, might further suggest qualitative differences in resource consumption between these two types of load (diagnosticity). Specifically, one would expect using fNIRS to study extraneous information that is not conducive to learning would predominantly show metabolic resource consumption of the posterior areas of the brain (association cortices) rather than more anterior areas like the PFC. 
\section{limitations}
\rev{There were limitations to the present work. First, there are likely individual differences not captured by our measures or models that may have influenced our findings.} As in many studies, the variability in these differences could dilute or skew the strength of the effects we saw. One such difference may lie in the sense of presence when in the VR setting which has been previously highlighted as impactful towards learning in VR by Grassini et al. \cite{ref19} and measured via neuroimaging in a study by Ronca et al. \cite{ref45}. To account for potential individual differences, the sample was balanced for biological sex and prior VR experience. While these variables were not the primary focus of the analysis, this balancing was intended to distribute their potential influence across the experimental conditions. We also acknowledge that there may have been some additional movement related noise in the fNIRS signal due to the naturalistic task that was not caught with our preprocessing methods.
\section{future work}
\rev{Adaptive training paradigms for complex, multi-step learning protocols could utilize these findings by implementing dynamic calibration of intrinsic load to optimally engage executive and multisensory integration processes based on these activation patterns.   Additionally, adaptive systems could monitor the right posterior parietal areas for activation if there is concern that extraneous load is impacting learning. Future research could further investigate the interplay of these cognitive resources and associated fNIRS markers in the adaptive training context by integrating EEG. This would allow for exploring how the differing temporal and spatial sensitivities of EEG and fNIRS contribute to responses to intrinsic load. Specifically, one such study could observe if fNIRS or EEG is more sensitive to different parts of a training paradigm (e.g., slower training phase compared to the rapid retention testing phase).}
\section{\rev{conclusion}}
This study contributes to the growing effort to make adaptive training simulators that use real-time brain data by providing insight into the relationships between the available brain signals and associated learning constructs. Notably, we found consistency with the literature regarding fNIRS-indicated brain activation across the types of workloads in the PFC and posterior parietal regions. \rev{Future work could seamlessly follow by implementing our paradigm and streaming the data to a machine learning algorithm that could intelligently adjust the training instructions or scaffolding structure.}

\bibliographystyle{IEEEtran}
\bibliography{references.bib}

\end{document}